\documentstyle[12pt]{article}
\begin{document}
\thispagestyle{empty}

\begin{center}
\LARGE \tt \bf {Neutrino asymmetry in general relativistic rotating radiative stars}
\end{center}

\vspace{2.5cm}

\begin{center} {\large 

L.C. Garcia de Andrade\footnote{Departamento de
F\'{\i}sica Te\'{o}rica - Instituto de F\'{\i}sica - UERJ

Rua S\~{a}o Fco. Xavier 524, Rio de Janeiro, RJ

Maracan\~{a}, CEP:20550-003 , Brasil.E-mail:garcia@dft.if.uerj.br}}
\end{center}
\vspace{2cm}

\begin{abstract}
Neutrino asymmetry in general relativistic radiative spacetime exterior to spinning stars is investigating by making use of Newmann-Penrose (NP) spin coefficient formalism. It is shown that neutrino current depends on the direction of rotation of the star. The solution is obtained in test field approximation where the neutrinos do not generate gravitational fields.
\end{abstract}

\newpage

 Previously A. Vilenkin \cite{1} has investigated the neutrino asymmetric around static relativistic stars. Earlier I. D. Soares \cite{2} have shown by making use of four component spinor fields that microscopic neutrino asymmetries , such as CPT violation exists in Vaidya's radiant background spacetime representing relativistic non-rotating radiative general relativistic (GR) stars. In this brief report we show that neutrino asymmetries can be applied to more general spacetimes such as radiative rotating spacetimes. More specifically we investigate neutrino asymmetries in the Kerr radiating spacetime representing the exterior of GR rotating relativistic stars. This solution of Einstein's field equations was discovered by Vaidya and Patel \cite{3} in 1973. The importance of investigating neutrino asymmetries goes from astrophysics to cosmology where the direction of the rotation of the universe can be determined by this asymmetry. Another application of the investigation of neutrinos in GR is on the gravitational colapse. This problem was investigated by Iyer, Dhurandhar and Vishveshwara \cite{4} which considered the massless Dirac equation in the background geometry relevant to the gravitational colapse, which means Friedmann dust interior matched by the Schwarzschild spacetime. Another important feature of this work is that no neutrinos are considered ouside black holes since as shown by J. Hartle \cite{5} there is no classical neutrinos outside black holes but only outside relativistic stable stars. Actually as shown by Wu and Cai et al \cite{6} only neutrinos in the form of Hawking radiation can be found outside black holes described by Kerr radiative metric. For further details on the NP formalism used here the reader is refereed to the review of Jogia and Griffiths (JG) \cite{7}. In this report we follow the notation of JG paper. Here $J^{\mu}$ is the neutrino current given by
\begin{equation}
J^{\mu}= {\phi}\bar{\phi} l^{\mu}
\end{equation}
where ${\phi}$ is the neutrino wave field which is obtained by making use of neutrino spinor
\begin{equation}
{\phi}^{A}={\phi} o^{A}
\end{equation}
where $o^{A}$ is the spinor dyad. This particular choice of spinor basis was chosen to simplify matters. This procedure allows us to write the neutrino equations 
\begin{equation}
D{\phi}=({\rho}- {\epsilon}){\phi}
\end{equation}
\begin{equation}
{\delta}{\phi}= ({\tau}- {\beta}){\phi}
\end{equation}
The only nonvanishing components of neutrino spin-coefficients are
\begin{equation}
{\mu}_{1}= -i{\phi}\bar{\phi}
\end{equation}
and 
\begin{equation}
{\gamma}_{1}= \frac{1}{2}i{\phi}\bar{\phi}
\end{equation}
\begin{equation}
{\Phi}_{00}= 0
\end{equation}
\begin{equation}
{\Phi}_{01}= -\frac{1}{2}ik{\phi}\bar{\phi}{\kappa}
\end{equation}
\begin{equation}
{\Phi}_{02}= -ik{\phi}\bar{\phi}{\sigma}
\end{equation}
\begin{equation}
{\Phi}_{11}= \frac{1}{2}ik{\phi}\bar{\phi}({\rho}-\bar{\rho})
\end{equation}
\begin{equation}
{\Phi}_{12}= \frac{1}{2}ik[{\phi}{\delta}\bar{\phi}+{\phi}\bar{\phi}(\bar{\alpha}-2{\tau})]
\end{equation}
\begin{equation}
{\Phi}_{22}=ik[{\phi}{\Delta}\bar{\phi}-\bar{\phi}{\Delta}{\phi}+{\phi}\bar{\phi}({\gamma}-\bar{\gamma})]
\end{equation}
where the ${\Phi}_{AB}$ , $A,B=0,1,2$, are the Ricci tensor NP components. Let us now considered the Kerr radiative metric \cite{3} in the coordinates $x^{0}=u$, $x^{1}=r$,$x^{2}=x$ and $x^{3}=y$ where the line element is given by
\begin{equation}
ds^{2}=(1-2mr{\rho}\bar{\rho})du^{2}+gdr+4mrasin^{2}x{\rho}\bar{\rho}dudy-{({\rho}\bar{\rho})}^{-1}dx^{2}-fdy^{2}
\end{equation}
where
\begin{equation}
g:=2(du-2asin^{2}xdy)
\end{equation}
and
\begin{equation}
f:= 2mra^{2}sin^{2}x {\rho}\bar{\rho}+ r^{2}+a^{2}sin^{2}x
\end{equation}
here $u$ is the retarded time coordinate and the speed of light in vacuum $c=1$. The expression $m(u)$ is the mass parameter. Besides a is a constant parameter like in the Kerr \cite{8} metric. where we have used the geometrical optics approximation up to order of $O(r^{-2})$ and drop out terms of order $O(r^{-3})$. This approximation allows us to obtain a very simple ghost neutrino solution. The spin coefficient ${\rho}$ appearing in the line element is given by 
\begin{equation}
{\rho}= -\frac{1}{(r-iacosx)}
\end{equation}
The null tetrad for this metric is computed by making use of the variables
\begin{equation}
{\Omega}=r^{2}+a^{2}
\end{equation}
and 
\begin{equation}
\Upsilon=\frac{r^{2}+a^{2}-2m(u)r}{2}
\end{equation}
The null tetrad for this metric is computed by making use of the variables
\begin{equation}
l_{\mu}={\delta}_{\mu}^{0}-a sin^{2}x {\delta}_{\mu}^{3}
\end{equation}
\begin{equation}
m_{\mu}= -\frac{\bar{\rho}}{\sqrt{2}}(iasinx{\delta}^{0}_{\mu}-({\rho}\bar{\rho})^{-1}{\delta}^{2}_{\mu}-i{\Omega}sinx{\delta}^{3}_{\mu})
\end{equation}
\begin{equation}
n_{\mu}= {\rho}\bar{\rho}[\Upsilon{\delta}_{\mu}^{0}-({\rho}\bar{\rho})^{-1}{\delta}^{1}_{\mu}-asin^{2}x{\delta}_{\mu}^{3}\Upsilon]
\end{equation}
From this tetrad we can obtain the following  differential operators
\begin{equation}
D=\frac{{\partial}}{{\partial}r}
\end{equation}
\begin{equation}
{\delta}=-\frac{\bar{\rho}}{\sqrt{2}}(iasinx\frac{{\partial}}{{\partial}u}+\frac{{\partial}}{{\partial}x}+icosecx\frac{{\partial}}{{\partial}y})
\end{equation}
\begin{equation}
{\Delta}={\bar{\rho}}{\rho}[-\Upsilon\frac{{\partial}}{{\partial}r}+{\Omega}\frac{{\partial}}{{\partial}u}+a\frac{{\partial}}{{\partial}y}]
\end{equation}
From this tetrad one is able to show that the following spin-coefficients vanish
\begin{equation}
{\epsilon}^{0}={\lambda}^{0}={\sigma}^{0}={\kappa}^{0}=0
\end{equation}
and 
\begin{equation}
{\pi}= \frac{iasinx {\rho}^{2}}{\sqrt{2}}
\end{equation}
\begin{equation}
{\beta}= -\frac{cotx \bar{\rho}}{2\sqrt{2}}
\end{equation}
\begin{equation}
{\alpha}= {\pi}-\bar{\beta}
\end{equation}
\begin{equation}
{\mu}= \Upsilon{\rho}^{2}\bar{\rho}
\end{equation}
\begin{equation}
{\nu}=-i{\dot{\bar{m}}}ra \frac{sinx {\rho}^{2}\bar{\rho}}{\sqrt{2}}
\end{equation}
\begin{equation}
{\gamma}= {\mu}+[r-m(u)]\frac{{\rho}\bar{\rho}}{\sqrt{2}}
\end{equation}
\begin{equation}
{\tau}=-ia \frac{sinx {\rho}^{2}\bar{\rho}}{\sqrt{2}}
\end{equation}
The radiative Kerr metric \cite{3} reduces to the Vaidya metric when the angular momentum of the compact object such as a black hole vanishes. Let us now use the method of separation of variables to solve the neutrino equations. Before that we note that the neutrino equations can be reduced to a simpler form by making use of the spin coefficients that vanish for this Kerr radiative metric. Thus
\begin{equation}
D{\phi}={\rho}{\phi}
\end{equation}
\begin{equation}
{\delta}{\phi}= ({\tau}-{\beta}){\phi}
\end{equation}
by performing the separation of variables ${\phi}(u,r,x,y)=A(u)B(r)C(x)$ where we have considered the following symmetry to simplify matters  ${\cal L}_{\frac{{\partial}}{{\partial}y}}{\phi}=\frac{\partial}{{\partial}y}{\phi}=0$ , where ${\cal L}$ represents the Lie derivative. Besides this simplification we consider the solution to be valid in spacetimes regions where $r>>a$ or far away from the compact stellar object we are considering. Applying this weak field approximation on the NP spin coefficients we obtain ${\rho}=\frac{-1}{r}$ and ${\tau}$ can be made to vanish. These simplifications imply that the neutrino PDE reduce to three ODE equations 
\begin{equation}
ia\frac{dA}{du}=nA
\end{equation}
\begin{equation}
\frac{dB}{dr}= -\frac{1}{r}B
\end{equation}
\begin{equation}
\frac{dC}{dx}+[nasinx+\frac{cotx}{2}]C=0
\end{equation}
Here n represents the constant parameter of the separation of variables of the PDE. The easiest equation to solve is the first one which yields
\begin{equation}
A(u)= e^{-inu}
\end{equation}
The second equation could be easily solved to yield 
\begin{equation}
B(r)= \frac{1}{r}
\end{equation}
Finally the solution of the last equation can now be solved, and $C(x)$ reads
\begin{equation}
C(x)=\frac{e^{nacosx}}{{sin}^{\frac{1}{2}}x}
\end{equation}
Collecting the three solutions of the ODEs one obtains the neutrino solution in the form
\begin{equation}
{\phi}(u,r,x)= \frac{e^{na(cosx-i\frac{u}{a})}}{r {sin}^{\frac{1}{2}}x}
\end{equation}
This solution represents physically a progressive neutrino wave propagating on the field of a rotating radiating star such as a spinning neutron star. At the plane $x=\frac{\pi}{2}$ this neutrino wave function reduces to 
\begin{equation}
{\phi}(u,r,x)= \frac{e^{-inu}}{r}
\end{equation}
This solution represents a spherically symmetric neutrino wave. Note that since the cosx function is bound so is the neutrino solution. By considering that a constant two-component neutrino spinor field is a solution of the neutrino equation for the Kerr radiative spacetime and due to the spin-coefficients of the Kerr radiative metric one obtains
\begin{equation}
D{\phi}=0
\end{equation}
\begin{equation}
{\delta}{\phi}=0
\end{equation}
which implies the following constraints on the spin-coefficients ${\rho}^{0}=0$ which contradict the computation of the spin-coefficients for that metric which allows us to conclude that there is no constant two-dimensional constant spinor neutrino field in Kerr radiative background as we wish to prove. Now let us compute the curvature and torsion scalars for this solution. This can be easily done if one notices that ${\phi}\bar{\phi}=\frac{e^{2nacosx}}{r^{2}sinx} $. Further simplification is obtained by making use the expressions for the torsion scalars
\begin{equation}
{\Theta}_{00}= {\Theta}_{01}= {\Theta}_{02}= {\Theta}_{11}={\Theta}_{22}=0
\end{equation}
\begin{equation}
i{\Theta}_{12}={\Phi}_{12}
\end{equation}
while Ricci curvature expressions are given by
where
\begin{equation}
{\Phi}_{12}= \frac{na e^{2nacosx}}{r^{2}}
\end{equation}
and
\begin{equation}
{\Phi}_{22}= \frac{ik}{r^{3}sinx}[Y-in{\Omega}] e^{2nacosx}
\end{equation}
The leptonic neutrino current 
\begin{equation}
J^{\mu}={\phi}\bar{\phi}l^{\mu}= \frac{e^{2nacosx}}{r^{2}sinx} l^{\mu}
\end{equation}
A simple observation of the RHS of this expression leads us to note taht when the angular momentum per unit mass is negative  $(a<0)$ the leptonic torsioned current is weaker than when the parameter $(a>0)$. Neutrinos in radiative Kerr background are presented as a solution of Weyl equation with Kerr radiative null tetrads. Chandrasekhar \cite{9} succeeded in finding a neutrino wave solution in Kerr background but it seems that only numerical solutions are possible to the exact problem including spinor solutions which also depends on the coordinate y. Neutrino test fields have been used in other backgrounds \cite{10} such as the G\"{o}del cosmological model to investigate the asymetries between neutrinos and antineutrinos. Modern aspects of massless neutrinos involve the investigation of geometrical optics of neutrinos in torsion potentials such as domain walls \cite{11} and other torsion potential \cite{12} in the strong gravity regime. In the context of massive neutrinos where it is responsible for neutrino oscillations \cite{13} but also may be an important mechanism in massless neutrinos case. 
\section*{Acknowledgements}
\paragraph*{}
I am very much indebt to Professors I.D. Soares and P.S.Letelier for helpful discussions on the subject of this paper. Special thanks go to Professor Jerry Griffiths for his kind attention and interest in our work. Thanks are also due to Professor Paul Tod for suggesting me this line of work many years ago during my post-doctoral at the Mathematical Institute of Oxford University. Grants from CNPq (Ministry of Science of Brazilian Government) and Universidade do Estado do Rio de Janeiro (UERJ) are acknowledged.

\end{document}